\documentclass{ws-ijmpd}
\usepackage[super,compress]{cite}
\usepackage{graphicx}
\begin{document}

\markboth{A. Z. Dolginov}
{THE RIEMANN GEOMETRY OF SPACE AND GRAVITATIONAL WAVES WITH THE SPIN $S=1$}

\catchline{}{}{}{}{}
%
%%%%%%%%%%%%%%%%%%%%%%%%%%%%%%%%%%%%%%%%%%%%%%%%%%%%%%%%%%%%%%%%%%%%

\title{THE RIEMANN GEOMETRY OF SPACE AND GRAVITATIONAL WAVES WITH THE SPIN $S=1$}

\author{ARKADY Z. DOLGINOV}

\address{Rensselaer Polytechnic Institute, \\
Troy, NY 12180, USA\\
arkady.dolginov@gmail.com\\dolgia@rpi.edu}

\maketitle

\begin{history}
\received{14/05/2014}
%\revised{Day Month Year}
\end{history}
\begin{abstract}
It is taken into account that not the Ricci tensor  ${R_{il}}$ (Einstein equation), but the Riemann tensor ${R_{iklm}}$  provides the most general description of the space geometry.  If ${{R_{il}}=0}$  (the space empty with matter, but it can be occupied by gravitational waves) then ${{R_{iklm}}={C_{iklm}}} $ . The tensor ${C_{iklm}}$  is the Weyl tensor, which disappears by conversion:${{R_{il}}={g^{km}}{R_{iklm}}}$ and we lose all information about the space structure, which is described by ${C_{iklm}}$ .The symmetry of  ${R_{il}}$   provides the existents of gravitational waves with the spin s=2. We show that ${C_{iklm}}$  describes gravitational waves with s=1. Such gravitational waves can be created in inhomogeneous media, where the selected directions are determined by derivates of the energy-momentum tensor ${T^m}_{i,k}$ of matter. It is taken into account that gravitation is described not only by the metric tensor
 $g^{ik} = (1/2)(\gamma^i \gamma^k + \gamma^k \gamma^i)$,  but also by the anti-symmetric tensor $\sigma^{ik} = (i/2)({\gamma^i}{\gamma^k} - {\gamma^k}{\gamma^i})$ , where  ${\gamma^k}(x)$ are Clifford matrices. 
 We show that the tensor ${K_{ik}} = (1/4){\sigma^{lm}} {R_{iklm}}$, is the anty-simmetric analog to the Ricci tensor. The ${K_{lm}}$ describes various kinds of space metrics, not described by the Ricci tensor. It includes the Weyl tensor  ${C_{iklm}}$, because $\sigma^{lm} {C_{lmik}}$ is not zero.
\end{abstract}

\ccode{PACS numbers:}

\section{Introduction}
In this paper  we take into account that the most general description of the space geometry is presented by the Riemann tensor. The Ricci tensor and the scalar $R=g^{km}R_{km}$ contain less information. The Riemann tensor has more independent components than the Ricci tensor \cite{Landau75}. In particular, using the Ricci tensor, we lose the information on the Weyl tensor, which is the part of the Riemann tensor, but dissapears by conversion in the Ricci tensor because ${g^{il}}{C_{iklm}}=0$. Therefore, using only the Ricci tensor, we can lose some information which is related to anti-symmetric properties of the Riemann tensor and, hence, some information on possible gravitational structures. We shall show that the tensor  ${K_{lm}} = (1/4){\sigma ^{ik}}{R_{iklm}}$,  where  ${\sigma^{ik}}(x)$ is the spin matrix and ${R_{iklm}}$ is the Riemann tensor, describes  gravitational structures  of space, different to that described by the Ricci tensor $R_{km} = g^{il} R_{iklm}$.   Tensors  ${K_{lm}}$ and ${R_{il}}$  are expressed by different combination of the ${R_{iklm}}$ components. These tensors depend on different components of energy-momentum tensors. They determine different sets of tetrads ${h^k}_\alpha $ and, hence, describe different cases of the space geometry. The  ${R_{il}} = 0$ does not mean that ${K_{lm}}$  and ${R_{iklm}}$  are zero.  The tensor ${K_{lm}}$ was considered in literature \cite{Pagels1965,Chapman1984},   as a tensor which describes only the spin properties of test particles, but not as the tensor which describes gravitational fields. We take into account that the spin matrix  ${\sigma^{ik}}(x)$, refers not only to the spin of a test particle but, as well as the metric tensor ${g^{km}}(x)$, determines the geometry of space. We shall show that the tensor ${K_{lm}}$  can be considered as the tensor complementary to the Ricci tensor. The Einstein equation describes gravitational waves with the spin s=2. It will be shown below that the tensor ${K_{lm}}$, which includes the Weyl tensor, describes gravitational waves with the spin s=1. Such waves can be created in media if there are selected directions, determined by non uniform distribution of matter.  
   To provide the independence of the final results from the chosen representation of the matrix ${\sigma^{ik}}(x)$ , the traces of quadratic forms are used. For example, the scalar curvature can be presented as: $R={g^{km}R_{km}}={Trace}[{\sigma^{ik}}{K_{ik}}]$.

\section{The tensor ${K_{lm}}$}	

   The most general expression of the Riemann tensor which is: (a) linear with respect to energy momentum tensors $T_{km}$, (b) for which the corresponding Ricci tensor determines Einstein equation, and  (c) which is not zero for the non flat geometry of the empty space, has the form\cite{Landau75}:
\begin{equation}\label{eq1}
\begin{array}{l}
{R_{iklm}} = \frac{1}{2}\{ \frac{{{\partial ^2}{g_{im}}}}{{\partial {x^k}\partial {x^l}}} + \frac{{{\partial ^2}{g_{_{kl}}}}}{{\partial {x^i}\partial {x^m}}} - \frac{{{\partial ^2}{g_{il}}}}{{\partial {x^k}\partial {x^m}}} - \frac{{{\partial ^2}{g_{km}}}}{{\partial {x^i}\partial {x^l}}} \} +
{g_{np}}\{ {\Gamma ^n}_{kl}{\Gamma ^p}_{im} - {\Gamma ^n}_{km}{\Gamma ^p}_{il}\}  \\ \\ = 
\frac{{4\pi G}}{{{c^4}}}\{ {g_{km}}{T_{il}} - {g_{im}}{T_{lk}} 
+ 
{g_{il}}{T_{km}} - {g_{kl}}{T_{im}} - \frac{2}{3}({g_{il}}{g_{km}} - {g_{kl}}{g_{im}})T\} + {({R_{iklm}})_0}\\ \\=

\frac{1}{2}\{{R_{il}}{g_{km}}-{R_{im}}{g_{kl}}+{R_{kl}}{g_{im}}-{R_{km}}{g_{il}}-\frac{1}{3}R({g_{il}}{g_{km}}-{g_{im}}{g_{kl}})\}+
{C_{iklm}}
\end{array}
\end{equation}
Here $(R_{iklm})_0$ is the part of the Riemann tensor which describes the geometry of the empty space. 
\begin{equation}\label{eq2}
\begin{array}{l}
{({R_{iklm}})_0} = \frac{1}{3}({g_{kl}}{g_{im}} - {g_{il}}{g_{km}}\} \Lambda + {C_{iklm}}
\end{array}
\end{equation}
Here $\Lambda $ is, so called, cosmological constant and $C_{iklm}$ is the Weyl tensor\cite{Weyl1927}.
The Weyl tensor does not depend on the energy-momentum tensors of the matter. It describes the geometry of empty space.
The Ricci tensor does not depend on the Weil tensor because ${{g^{il}}{C_{iklm}}=0}$,  however
 ${\sigma^{ik}}{C_{iklm}}$ is not zero if the geometry of the empty space is not Euclidian. The physical vacuum is a complicated system and it is possible that the vacuum is inhomogeneous at cosmical scales\cite{Dolginov 2011}. 
Using the expression (\ref{eq1}) we obtain the Ricci second rank tensor as:
\begin{equation}\label{eq3}
\begin{array}{l}
{R_{km}} = {g^{il}}{R_{iklm}} = \frac{{8\pi G}}{{{c^4}}}\{ {T_{km}} - \frac{1}{2}{g_{km}}T\} + {({R_{km}})_0}, \\  \\
{({R_{km}})_0} = {g^{il}}{({R_{iklm}})_0} = - {g_{km}}\Lambda 
\end{array}
\end{equation}
The metric tensor ${g^{km}}(x)$ and the spin tensor matrix ${\sigma^{ik}}(x)$ are determined by the Clifford matrices ${\gamma^k}(x)$:
\begin{equation}\label{eq4}\\
\begin{array}{l}
{g^{ik}} = (1/2)({\gamma ^i}{\gamma ^k} + {\gamma ^k}{\gamma ^i}),\,\,\,
{\sigma ^{ik}} = (i/2)({\gamma ^i}{\gamma ^k} - {\gamma ^k}{\gamma ^i}) \\ \\
{\gamma^k}(x) = h_\alpha^k(x){\gamma ^\alpha },\,\,\,
{g^{km}}(x) = h_\alpha ^k(x)h_\alpha ^m(x),\,\,\,\,\,
{\sigma ^{km}}(x) = h_\alpha ^k(x)h_\beta ^m(x){\sigma ^{\alpha \beta }}
\end{array}
\end{equation}
Clifford matrices ${\gamma ^k}(x)$ depend on space-time coordinates. Matrices ${\gamma^\alpha }$ are Clifford matrices for a flat space-time and $h_\alpha^k(x)$ are tetrads which determine the metric tensor ${g^{km}}(x)$ and the spin tensor ${\sigma ^{ik}}(x)$.
Latin indices ($n, km$, etc), were used here for matrices (${\gamma ^n},{\sigma ^{km}}$ , etc.) in curved space. Greek indices ($\alpha ,\alpha \beta$, etc.) determine (${\gamma ^\alpha }$, ${\sigma ^{\alpha \beta }}$, etc.) for the flat geometry. Matrices ${\gamma ^k}(x)$, ${\sigma ^{ik}}(x)$, etc. are used in quantum theory but, by themselves, they are classical quantities which are determined by the space geometry. The  ${\gamma ^k}(x)$ determines ${g^{km}}(x)$ and ${\sigma ^{ik}}(x)$, but both ${g^{km}}(x)$ and  ${\sigma ^{ik}}(x)$ are needed to determine  ${\gamma ^k}(x)$.
For example, the ${\gamma ^k}(x)$, but not  ${g^{km}}(x)$, determine gravitational effects in the expression for the current  created by electrons: $j_n = \sqrt { - g} {\psi ^ * }{\gamma _n}\psi$ and
${\nabla _n}{j^n} = 0$.

  The interval in the Riemann space can be presented as ${ds={\gamma_p}dx^p}$. In the case of two infinite-small displacements ${{dx_1}^n}$ and ${{dx_2}^m}$ at the point $x$, the associative intervals ${ds_1}$ and ${ds_2}$ do not commutate:
\begin{equation}\label{eq5}
i[d{s_1},d{s_2}] = {\sigma _{mn}}(dx_1^mdx_2^n - dx_2^mdx_1^n)
\end{equation}
The ${(dx^i) \times (dx^k)=- (dx^k) \times (dx^i)}$ is the infinitesimal element of surface area, determined by the vector displacements.
  Final results of the theory that can be compared with observations have to be presented in a form which is independent of the ${\gamma ^k}(x)$ representation. For this purpose quadratic forms and traces of the resulting expressions are taken as final results. The quantity ${ds^2 = g_{ik}{dx^i}{dx^k}}$ can be measured with arbitrary precision with rods and clocks. It is similar to the situation in quantum theory, where the wave function is dependent on the ${\gamma ^k}(x)$ representation, but only the square of this function describes observations.
 
  To consider the problem of gravitational waves with the spin $s=1$, it is necessary to use some properties of spinors and  properties of  the anti-symmetric second rank tensor ${K_{ik}} $. Not only scalars, vectors and tensors, but also spinors and spin-tensors, are classical geometrical functions, which are representing the group of rotation. Spinors and Clifford matrices do not contain the Plank constant and can be used to describe  classical as well as quantum systems. The covariant derivative ${\nabla_n}$ of a spinor $\psi $ is \cite{Fock1929}: 
\begin{equation}\label{eq6}
\begin{array}{l} 
{\nabla _n}\psi = {\psi _{,n}} + {i}{\Gamma _n}\psi \\ \\
{\Gamma _n} = (1/4){\gamma ^m}{\gamma _{m;n}} = \frac{1}{4}\{ h_{(\alpha )}^m\frac{{d{h_{m(\beta )}}}}{{d{x_n}}} - h_{(\alpha )}^m{h_{l(\beta )}}\Gamma _{mn}^l\} {\sigma ^{(\alpha \beta )}} \\ \\
{\nabla _m}{\gamma ^n} = {\gamma ^n}_{;m} +{i}{\Gamma _m}{\gamma ^n} -{i}{\gamma ^n}{\Gamma _m} = 0 \\ \\
0 = {\nabla _\beta }({\nabla _\alpha }{\gamma _\mu }) - {\nabla _\alpha }({\nabla _\beta }{\gamma _\mu }) = {R_{\mu \lambda \alpha \beta }}{\gamma ^\lambda } + [{K_{\alpha \beta }},{\gamma _\mu }]
\end{array}
\end{equation}
Using expressions (\ref{eq6}) for the gamma matrix covariant derivatives, the tensor ${K_{ik}}$ can be expressed in terms of Christoffel symbols ${\Gamma_n}$ for spinors as \cite{Pagels1965}: 
\begin{equation}\label{eq7}
{K_{ik}} = \frac{1}{4}{\sigma ^{lm}}{R_{iklm}} = \frac{{\partial {\Gamma _i}}}{{\partial {x^k}}} - \frac{{\partial {\Gamma _k}}}{{\partial {x^i}}} + {i}{\Gamma _k}{\Gamma _i} -{i}{\Gamma _i}{\Gamma _k}
\end{equation} 

Using expressions (\ref{eq1}) and (\ref{eq7}) the tensor ${K_{lm}}$ can be presented as \cite{Dolginov2011}:
\begin{equation}\label{eq8}
\begin{array}{l}
{K_{lm}} = {\rm{ }}\frac{{2\pi G}}{{{c^4}}}\{ {\sigma_{nm}}T^n_l - {\sigma_{nl}}T^n_m - \frac{2}{3}{\sigma _{lm}}T\}\,+\frac{1}{4}{({\sigma ^{ik}}{R_{iklm}})_0}
\end{array}
\end{equation}
Equations (\ref{eq1}-\ref{eq3}) and (\ref{eq7}-\ref{eq8}) are nonlinear. Only a few exact solutions of 
Einstein equation (\ref{eq3}) are known. Gravitational waves with the spin $s=2$ are described in literature  as a solution of equation ${R_{il}} = 0$. 
The equation for ${K_{lm}}$ can be obtained \cite{Pagels1965,Chapman1984} using the Bianchi identities. It is important to note that these equations have the form similar to Maxwell equations:  
\begin{equation}\label{eq9}
{\nabla _l}{K_{ik}}{\rm{ + }}{\nabla _i}{K_{kl}}{\rm{ + }}{\nabla _k}{K_{li}} = 0,\,\,\,\,\,\,\,\,\,\,\,{\rm{ }}{\nabla _n}{K^{nm}} = - {\rm{ }}{J^m}
\end{equation}
Here ${J^m}$ determines the divergence-free vector, similar to the electric current. 
\begin{equation}\label{eq10}
{J^m} = \frac{{2\pi G}}{{{c^4}}}{\sigma ^{ik}}T^m_{i,k} + {({J^m})_0},\,\,\,\,\,\,\,\,\,\, {\nabla _m}{J^m} = 0
\end{equation}
It was taken into account that ,according to equation (6) : ${\nabla_n}{g^{il}}={\nabla_n}{\sigma^{ik}}=0$.
The expression for ${J^m}$ was obtained using the equation (1) and the contracted Bianchi identities in the form:  ${R^{iklm}_{,l}}={R^{i,k}_{m}}-{R^{k,i}_{m}}$. The direction of the current ${J^m}$ is determined by the derivatives of the energy-momentum tensor ${T^m_{I,k}}$. In a particular case of continuous media: ${T^m_n}=(p+\epsilon){u_m}{u^n}-p(\delta^m_n)$ and $T_n^n = \varepsilon  - 3p$ where $\varepsilon$ is the energy density,  $p$ is the pressure and ${u^n}$ is the 4-velocity of the media.

The theory allows adding some polar vector to ${\Gamma _n}$. This vector can be interpreted
as a vector potential of electromagnetic field and, hence, can introduce this field
into the theory.  It is seen from equation (\ref{eq8}) that the tensor ${K_{ik}}$ has a potential and vortex part, similar to the electromagnetic field ${F_{ik}}$. 
\section{Gravitational waves with the spin  $s=1$}
There are well known statements that gravitational waves with the spin $s=1$ cannot exists, because gravitational forces are attractive and are determined by tensor quantities, in contrast to electromagnetic processes, which are determined by vectors \cite{Thorne1980,Landau75,Dyson2013}. Indead, the symmetry of the tensor Ricci excludes the possibility to describe gravitational waves with the spin different from s=2. However, it does not contradicts to the symmetry of the  tensor ${K_{ik}}$ . The existence of the selected directions, which are connected with locations of positive and negative charges, determines the directions of electric currents in space.  It provides the existence of electromagnetic waves with the spin $s=1$.  The selected directions can exists in gravitational fields if the distribution of matter, which creates gravitation, is non-uniform.  We will show that in this case the emission of gravitational waves with the spin $s=1$ is possible. The current, which create these waves, is determined by derivatives of the energy-momentum tensor ${T^m}_{i,k}$ of  the matter and by the Clifford matrices ${\gamma^k}(x)$ .     
      Using the analogy of equations (\ref{eq9}) for $K_{lm}$ and Maxwell equations for the electromagnetic field $F_{lm}$, and introducing quantities ${\Phi _k} = {{\rm{K}}_{0k}} + {\rm{i}}{{\rm{K}}_{{\rm{lm}}}}$ ,where $k\ne l \ne  m \ne 0$, and ${\Theta _k} = {\rm{ }}{{\rm{K}}_{0k}} - {\rm{i}}{{\rm{K}}_{{\rm{lm}}}}$ , we can obtain the equations for ${\Phi _k}$ and ${\Theta _k}$:
\begin{equation}\label{eq11}
\begin{array}{l}
i\frac{\partial }{{\partial t}}\left[ \begin{array}{l}
{\Phi _1}\\
{\Phi _2}\\
{\Phi _3} \\
\end{array} \right]=\left[ \begin{array}{lll}
0 & 0 & 0\\
0 & 0 & - i\frac{\partial }{{\partial x}}\\
0 & i\frac{\partial }{{\partial x}} & 0 \\
\end{array} \right]\left[ \begin{array}{l}
{\Phi_1}\\
{\Phi_2}\\
{\Phi_3} \\
\end{array} \right] 
+ 
\left[ \begin{array}{lll}
0 & 0 & i\frac{\partial }{{\partial y}}\\
0 &0 &0\\
- i\frac{\partial }{{\partial y}}& 0& 0 \\
\end{array} \right]\left[ \begin{array}{l}
{\Phi _1}\\
{\Phi _2}\\
{\Phi _3} \\
\end{array} \right] + \left[ \begin{array}{lll}
0 & - i\frac{\partial }{{\partial z}} & 0\\
i\frac{\partial }{{\partial z}} & 0 & 0\\
0 & 0 & 0 \\
\end{array} \right]\left[ \begin{array}{l}
{\Phi _1}\\
{\Phi _2}\\
{\Phi _3} \\
\end{array} \right]\\
- i\frac{\partial }{{\partial t}}\left[ \begin{array}{l}
{\Theta _1}\\
{\Theta _2}\\
{\Theta _3} \\
\end{array} \right] = \left[ \begin{array}{lll}
0 & 0 &0\\
0 & 0 & - i\frac{\partial }{{\partial x}}\\
0& i\frac{\partial }{{\partial x}} &0 \\ 
\end{array} \right]\left[ \begin{array}{l}
{\Theta _1}\\
{\Theta _2}\\
{\Theta _3} \\
\end{array} \right] 
+ 
\left[ \begin{array}{lll}
0&0&i\frac{\partial }{{\partial y}}\\
0&0&0\\
- i\frac{\partial }{{\partial y}}&0&0
\end{array} \right]\left[ \begin{array}{l}
{\Theta _1}\\
{\Theta _2}\\
{\Theta _3}
\end{array} \right] +  \left[ \begin{array}{lll}
0 &- i\frac{\partial }{{\partial z}} & 0\\
i\frac{\partial }{{\partial z}} & 0 & 0\\
0 &0 & 0 \\
\end{array} \right]\left[ \begin{array}{l}
{\Theta _1}\\
{\Theta _2}\\
{\Theta _3} \\ 
\end{array} \right]
\end{array} 
\end{equation}
Here $\Phi_1 = K_{01} + iK_{23}$, $\Phi_2 = K_{02} + iK_{31}$, etc., and 
$\Theta_1 = K_{01} - iK_{23}$, etc. The units $c=1$,$\hbar=1$ are used here. 
The equations (\ref{eq11}) can be written in a form:
\begin{equation}\label{eq12}
\begin{array}{l}
i\frac{{\partial \Phi }}{{\partial t}} = {\bf{sp}}\Phi, \,\,\,\, {\bf{p}}\Phi = 0, \,\,\,\,
i\frac{{\partial \Theta }}{{\partial t}} = - {\bf{sp}}\Theta, \\ \\
{\bf{p}}\Theta = 0, \,\,\,\,
{\bf{p}} = \nabla 
\end{array} 
\end{equation}
\begin{equation}\nonumber
\begin{array}{l}
{s_x} = \left[ \begin{array}{lll}
0&0&0\\
0&0 &- i\\
0&i&0 \\
\end{array} \right],{s_y} = \left[ \begin{array}{lll}
0&0&i\\
0&0&0\\
- i&0&0\\
\end{array} \right], 
{s_z} = \left[ \begin{array}{lll}
0 &- i&0\\
i&0&0\\
0&0&0
\end{array} \right];
{s^2} = \left[ \begin{array}{lll}
2&0&0\\
0&2&0\\
0&0&2 \\
\end{array} \right];
\\ \\
\Phi = \left[ \begin{array}{l}
{\Phi _1}\\
{\Phi _2}\\
{\Phi _3}
\end{array} \right];\Theta = \left[ \begin{array}{l}
{\Theta _1}\\
{\Theta _2}\\
{\Theta _3}
\end{array} \right], \,\,\,
\Psi = \left( \begin{array}{l}
\Phi \\
\Theta 
\end{array} \right)
\end{array}
\end{equation}
%\end{comment}
%
%
The matrix $s$ has the properties ${s_i}{s_k} - {s_k}s{}_i = i{\varepsilon _{ikl}}{s_l}$ and 
${s^2} = s(s + 1)$, which means that $s = 1$. Therefore, $s$ can be considered as the spin of a particle which is described by the wave function $\Psi$.

In the general case the $\Phi$ and $\Theta$ are complicated functions of space and time. These functions can be presented as a sum of plain waves. In a simplest case
\begin{equation}\label{eq13}
\Phi = {\Phi _0}{e^{i{\bf k}\cdot{\bf r} - i\omega t}}, \,\,\,\, \Theta = {\Theta _0}{e^{i{\bf k}\cdot{\bf r} - i\omega t}}
\end{equation} 
Here the $\Phi$ presents a wave with the spirality +1 and $\Theta$ presents a wave with the spirality -1. 
\begin{equation}\label{eq14}                
 {\bf (k\cdot s)}{\Theta_k} =  - {\omega_k}{\Theta_k},\,\,\,\, 
{\bf(k\cdot s)}{\Phi_k} = {\omega_k}{\Phi_k}
\end{equation}

Equations (\ref{eq12}) are similar to the Weyl equations \cite{Weyl1929} for neutrino with the zero rest mass. For the neutrino the $\Phi$ and $\Theta$ are components of bispinor and ${\bf{s}} = \sigma /2$ . Reflection operation transform $\Theta \mathbin{\lower.3ex\hbox{$\buildrel\textstyle\rightarrow\over
{\smash{\leftarrow}\vphantom{_{\vbox to.5ex{\vss}}}}$}} \Phi $ as well as for the case of neutrino. 
To provide the independence of the final results from some specific representation of the matrix ${\sigma ^{ik}}$ , the quadratic forms should be used, and the trace of the final formulae should be considered. Such procedures are common in quantum theory but, by themselves, those are classical procedures.
\begin{equation}\label{eq15}
\begin{array}{l}
Tr\{ ({\sigma ^{ik}})({\sigma ^{lm}})\}  = 2[{g^{km}}{g^{il}} - {g^{im}}{g^{kl}}]
\end{array}
\end{equation}
\begin{equation}\label{eq16}
\left| \Psi  \right|^2 = Tr (\sum\limits_n \{  {\left| {{\Phi_n}} \right|^2} + {\left| {{\Theta_n}} \right|^2}  \} ) =
 \frac{1}{4} \sum\limits_{n,p,s}  \{  ({R^{ps}}_{0n} ){R_{pson}} +  ({R^{ps}}_{lm}){R_{pslm}} \} 
\end{equation}
The indexes $n,l,m$ are correspondingly: 123, 231 and 312. 
 
Results expressed in terms of ${\left| \Psi \right|^2}$ can be compared with observations. As in the case of electromagnetic waves, the wave function $\Psi $, determines the amplitude of the gravitational perturbation, but not the probability to find a quant of gravitation at some point of space. This function has a rather complicated dependence from the metric tensor ${g_{ik}}$. The waves with the spin $s = 1$, determined by ${\left| \Psi \right|^2}$, are not similar to gravitational waves with the spin $s = 2$, which are determined directly by ${g_{ik}}$ \cite{Landau75}. 

    It can be seen from the equations (9) and (10) that the source of the spin curvature is the current ${J_m}$ which is determined by changes of the energy momentum tensors ${T_m^{n,l}}$  of the surrounding matter in space and time. This current creates gravitational wave. The wave propagates and interacts with the matter around. The wave changes the state of the surrounding matter and creates the current ${J_m}$ at the point of observation. The ${J_m}$ determines ${K_{lm}}$  and, hence, the functions ${\Phi _k} = {{\rm{K}}_{0k}} + {\rm{i}}{{\rm{K}}_{{\rm{lm}}}}$ , and ${\Theta _k} = {\rm{ }}{{\rm{K}}_{0k}} - {\rm{i}}{{\rm{K}}_{{\rm{lm}}}}$, according expressions (9) and (10). The interaction of a wave with equipment, used for wave detection, changes the energy momentum tensor of the matter at the point of observation and can be detected.        
       Unfortunately, gravitational waves of any kind were not observed up to now. Investigations of complicated, time dependent changes of gravitational forces are necessary to detect the gravitational waves. The complicated technical problem of the gravitational waves detection needs a special consideration and is not considered here.

   According to the theory of the Universe development \cite{Linde1990}, gravitational perturbations, in particular in the form of gravitational waves, are created not only by motions of celestial bodies, but were also created  after the big bang. The distribution of such perturbations has to be inhomogeneous similar to distribution of the relict electromagnetic radiation.

\section{Conclusion}
1. For complete description of the space geometry it is necessary to use the tensor $R_{iklm}$.  The tensor ${R_{km}}$  is not sufficient for this description.  The tensor ${K_{ik}}$  can be considered as a tensor complimentary to ${R_{km}}$. The ${K_{ik}}$    describes space structures different to those described by ${R_{km}}$.

2. In particular, the equation for ${K_{ik}}$  describes gravitational waves with the spin $s=1$ .

Acknowlegements

 The author wishes to thank Professor I.N. Toptygin of the St.Petersburg Politechnical University and the Deputy Director of the Pulcovo observatory (Russia) Professor Yu.N. Gnedin for helpful discussions.

\bibliographystyle{ws-ijmpd}
\bibliography{dedabib}
%\begin{thebibliography}{0}   
%\end{thebibliography}

\end{document}